\documentclass[twocolumn]{aastex62}

\usepackage{xspace}
\usepackage{url}
\usepackage{longtable}
\usepackage{hyperref}
\usepackage{array}
\usepackage{float}
\usepackage{threeparttable}
\usepackage{url}

\newcommand{\io}[2]{#1\,{\textsc{#2}}}

\def\Lsun{{\rm L$_{\odot}$}\xspace}
\def\Rsun{{\rm R$_{\odot}$}\xspace}
\def\Msun{{\rm M$_{\odot}$}\xspace}

\def\halpha{{\rm H$\alpha$}\xspace}
\def\he2{{He~{\small II}}\xspace}

\graphicspath{{./}{figures/}}

\submitjournal{ApJ}

\shorttitle{V906 Car}
\shortauthors{Wee et al.}

\begin{document}

\title{Multi-wavelength Photometry and Progenitor Analysis of the Nova V906 Car}

\correspondingauthor{Jerrick Wee}
\email{weejerrick@u.yale-nus.edu.sg}

\author[0000-0002-8654-568X]{Jerrick Wee}
\affil{Yale-NUS College, 16 College Avenue West, Singapore 138527}

\author[0000-0003-0901-1606]{Nadejda Blagorodnova}
\affiliation{Department of Astrophysics/IMAPP, Radboud University, Nijmegen, The Netherlands}

\author{Bryan Edward Penprase}
\affiliation{Soka University of America, 1 University Drive, Aliso Viejo, CA 92656, USA}

\author{Jett Pierce Facey}
\affiliation{Soka University of America, 1 University Drive, Aliso Viejo, CA 92656, USA}

\author{Taiga Morioka}
\affiliation{Soka University of America, 1 University Drive, Aliso Viejo, CA 92656, USA}

\author[0000-0002-6339-6706]{Hank Corbett}
\affil{University of North Carolina at Chapel Hill, 120 E. Cameron Ave., Chapel Hill, NC 27514, USA}

\author[0000-0002-8558-4353]{Brad N. Barlow}
\affiliation{Department of Physics, High Point University, One University Parkway, High Point, NC 27268, USA}

\author[0000-0002-6540-1484]{Thomas Kupfer}
\affiliation{Kavli Institute for Theoretical Physics, University of California, Santa Barbara, CA 93106, USA}

\author[0000-0001-9380-6457]{Nicholas M. Law}
\affil{University of North Carolina at Chapel Hill, 120 E. Cameron Ave., Chapel Hill, NC 27514, USA}

\author[0000-0001-8791-7388]{Jeffrey K. Ratzloff}
\affil{University of North Carolina at Chapel Hill, 120 E. Cameron Ave., Chapel Hill, NC 27514, USA}

\author[0000-0002-0583-0949]{Ward S. Howard}
\affil{University of North Carolina at Chapel Hill, 120 E. Cameron Ave., Chapel Hill, NC 27514, USA}

\author[0000-0001-5083-8272]{Ramses Gonzalez Chavez}
\affil{University of North Carolina at Chapel Hill, 120 E. Cameron Ave., Chapel Hill, NC 27514, USA}

\author[0000-0001-9981-4909]{Amy Glazier}
\affil{University of North Carolina at Chapel Hill, 120 E. Cameron Ave., Chapel Hill, NC 27514, USA}

\author[0000-0001-9981-4909]{Alan Vasquez Soto}
\affil{University of North Carolina at Chapel Hill, 120 E. Cameron Ave., Chapel Hill, NC 27514, USA}

\author{Takashi Horiuchi}
\affil{Division of Science, National Astronomical Observatory of Japan, 2-21-1 Osawa, Mitaka, Tokyo 181-8588, Japan}

\begin{abstract}
We present optical and infrared photometry of the classical nova V906 Car, also known as Nova Car 2018 and ASASSN-18fv, discovered by ASASS-SN survey on 16.32 March 2018 UT (MJD 58193.0). The nova reached its maximum on MJD 58222.56 at $V_{\rm{max}} = 5.84 \pm 0.09$ mag and had decline times of $t_{2,V} = 26.2 $ d and $t_{3,V} = 33.0 $ d. The data from Evryscope shows that the nova had already brightened to $g'\simeq 13$\,mag five days before discovery, as compared to its quiescent magnitude of $g=$20.13$\pm$0.03. The extinction towards the nova, as derived from high resolution spectroscopy, shows an estimate consistent with foreground extinction to the Carina Nebula of $A_V = 1.11_{-0.39}^{+0.54}$. The light curve resembles a rare C (cusp) class nova with a steep decline slope of $\alpha=-3.94$ post cusp flare. From the lightcurve decline rate, we estimate the mass of white dwarf to be $M_{WD}$ = $ < 0.8$M\textsubscript{\(\odot\)}, consistent with $M_{WD}=0.71^{+0.23}_{-0.19}$ derived from modelling the accretion disk of the system in quiescence. The donor star is likely a K-M dwarf of 0.23-0.43\,\Msun, which is being heated by its companion.
\end{abstract}

\keywords{classical novae: individual (V906 Car) --- dust, extinction --- techniques: photometric}

\section{Introduction} \label{sec:intro}

Eruptions of classical novae (CNe) occur on the surface of a mass-accreting white dwarf (WD) from a hydrogen-rich brown dwarf, red dwarf, red giant companion, or a helium star in a close binary system \citep{Warner1995CAS}. When sufficient mass is accumulated to the point where the degenerate electron pressure at the base of the mass envelope exceeds a critical value, a thermonuclear runaway occurs \citep{2016PASP..128e1001S}. Most or all of the envelope is ejected, and the luminosity increases up to or even beyond the WD Eddington luminosity \citep{1981ApJ...243..926S}. The luminosity of a nova eruption is typically in excess of $10^5L\textsubscript{\(\odot\)}$, making CNe among the most luminous stars in the galaxy \citep{2016ApJS..227....1S}. 

Historically, due to the lack of coverage, few novae have been caught on their initial rise \citep{2008clno.book.....B}, and it was only possible to construct parts of the early rise from patrol images of the sky taken by amateur astronomers \citep[see for example V1500 Cyg]{Liller1975IAUC}. However, with increasingly better transient event detectors, such as the All-Sky Survey for SuperNovae \cite[ASASSN;][]{Shappee2014ApJ} and Evryscope \citep{2016SPIE.9906E..1ML} among others, more CNe have been caught on these cameras on their early rise, allowing astronomers to gather precious early-time data of novae outbursts. These observations are crucial as they provide insights on many previously poorly observed and understood phases of novae evolution \citep{2016ApJ...820..104H}.

CNe can be classified through various properties, such as their orbital periods \citep{1984ApJS...54..443P}, magnetic strength \citep{Warner1995CAS}, and spectra \citep{Williams1991ApJ}. Another popular method of classifying CNe is through their light curve shape. Based on a statistical analysis by \citet{Strope2010AJ}, a majority of CNe are S class CNe, having a smooth, power-law decline with no major fluctuation. However, there are many peculiar novae that do not conform to this type of evolution. One such type is the C class nova, which has a characteristic secondary maximum after the primary peak and appears to be a substantial superimposed brightening above the base level power-law decline \citep{Strope2010AJ}. 

With the increased prevalence of spectral data of CNe, astronomers have found that there is a statistical correlation between decline rate and the mass of the progenitor WD. \cite{HachisuKato2006ApJS} developed an optical and near-infrared lightcurve model of novae in which bremsstrahlung from optically thin ejecta dominates the continuum flux. They derived a ``universal decline law'' and introduced a ``timescaling factor'' to their light curve template. Using this timescaling factor, \cite{HachisuKato2006ApJS} are able to estimate various nova properties, such as the optically thick wind phase and the WD mass. 


In this paper, we present the results of our photometric follow-up campaign of the nova V906 Car (also known as ASASSN-18fv and Nova Car 2018) and its progenitor system. The nova was discovered \citep{Stanek2018ATel11454}  by All Sky Automated Survey for Supernovae \cite[ASAS-SN]{Shappee2014ApJ} on 2018 March 20.32 UT ($T_D=58197$ MJD), which we will adopt as our reference epoch. The transient was located at RA=10\textsuperscript{h}36\textsuperscript{m}15\textsuperscript{s}.43 and Dec=$-59^{\circ}$35$'$53.73$"$ J2000. This position is updated using the Gaia DR2 data \citep{GaiaDR2} available for the progenitor, corrected to J2000. The nova is located near the Galactic plane, in the same field of view as the Carina Nebula. After discovery, multiple optical and near-infrared spectroscopic observations confirmed V906 Car to be a classical nova a few days after its discovery \citep{2018ATel11460....1L, 2018ATel11468....1I, 2018ATel11506....1R}. The nova was also serendipitously observed by the BRIght Target Explorer (BRITE) Constellation \citep{Weiss2014}, while monitoring the nearby red giant star HD 92063. The lightcurve, which contains the full rise of the nova, shows that the outburst  started at about 6 days before the date of discovery \footnote{\url{https://www.utias-sfl.net/?p=3015}}.

In a recent study of Car V906, \citet{2020NatAs.tmp...79A} analyzed high cadence concurrent $\gamma$-ray and optical data taken during the nova outburst. The high correlation between the flaring emission in both wavelengths suggested that the bulk of the nova luminosity was shock-powered. This challenges the traditional picture, where the nova luminosity originates in sustained nuclear burning on the surface of a WD after the initial eruption. The study also found that the X-ray emission in Car V906 was highly attenuated, consistent with highly embedded shocks in the ejecta \citep{2018ATel11608....1N}. The shock-powered X-ray emission was reprocessed and emerged at longer wavelengths. 

Here we focus on the observational properties of V906 Car and provide the first analysis of the system in quiescense. We present our follow-up observations of the nova in Section \ref{sec:style}, spanning from 6 day before discovery, up to a year after. We estimate the extinction towards the nova in Section \ref{sec:high-res_spec} and describe the nova photometric evolution in Section \ref{sec:photevol}. We analyse the nova lightcurve and the progenitor system in Section \ref{sec:analysis}. In Section \ref{sec:discussion} we discuss the classification of the nova and estimate its WD mass based on decline rates. We summarise our conclusions in Section \ref{sec:conclusion}.

\section{Observations} \label{sec:style}

\subsection{CTIO 1.3m ANDICAM} \label{subsec:ctio1.3m}

We used the Yale SMARTS 1.3m Telescope at the Cerro Tololo Inter-American Observatory (CTIO) to obtain optical and near-infrared (near-IR) observations in $BVRIJHK$ bandpasses. The images were taken under the program YNUS-16A-0001 (PI B. Penprase) with the ANDICAM instrument, an imager permanently mounted on the 1.3\,m telescope that takes simultaneous optical and infrared data. The ANDICAM instrument uses the standard Johnson $BV$ filters, Kron-Cousins $RI$ filters, and standard CIT/CTIO $JHK$ filters. The optical field of view is 6.3$' \times 6.3'$, while the IR field of view is $2.34' \times 2.34'$. With the CCD readout in 2$\times$2 binning mode, the ANDICAM instrument gives a plate scale on the 1.3\,m telescope of 0.369$''$ pixel$^{-1}$ for optical imaging and 0.274$''$ pixel$^{-1}$ for near-IR imaging. Further information can be found on the ANDICAM instrument specification website.\footnote{http://www.astronomy.ohio-state.edu/ANDICAM/detectors.html} 

\begin{figure}
\plotone{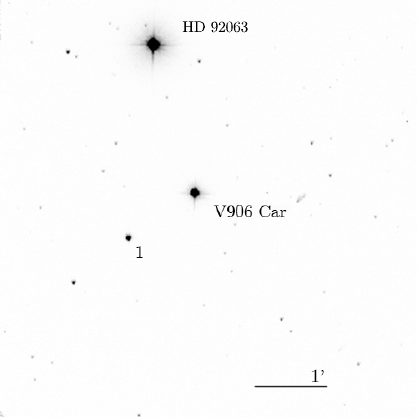}
\caption{$V$-band image of V906 Car off the Carina Nebula obtained with the CTIO 1.3m telescope on 2018 April 1. The exposure time was 5s. The local photometric standard is numbered. The bar corresponds to 1$'$. North is up, and east to the left. The nova is marked in the center of the image and the bright identifiable star HD 92063 north of the nova is labelled.\label{fig:imageop}}
\end{figure}

\begin{figure}
\plotone{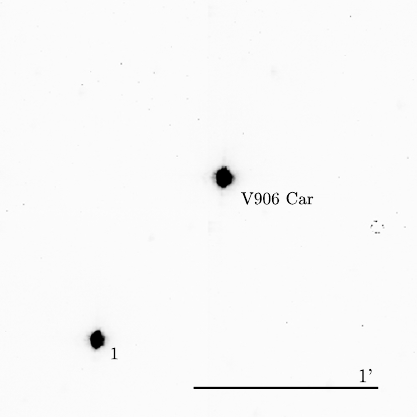}
\caption{$J$-band image of V906 Car off the Carina Nebula obtained with the CTIO 1.3m telescope on 2018 April 1. The exposure time was 5s. The local photometric standards are numbered. The bar corresponds to 1'. The nova is marked in the center of the image. North is up, east is left.\label{fig:imageir}}
\end{figure}

Our observations of the nova spanned 126 days, starting on 2018 March 25 and ending on 2018 July 29. We took daily cadence from 2018 March 25 to 2018 April 20. After that, we took weekly observations until 2019 July 29. Figure \ref{fig:imageop} shows a $V$-band image where we identify the location of the nova and the local field standard. Table \ref{tab:opseq} presents the optical photometry sequence of the local field. Figure \ref{fig:imageir} is a $J$-band image which shows the location of the nova and the field standards. The data on the local field standards for the near-IR can be found on Table \ref{tab:irseq}.

We conducted the photometric reduction of the optical data using the \texttt{Photutils}'s aperture photometry tool \citep{2016ascl.soft09011B}, a part of the \texttt{Python}-based package \texttt{Astropy} \citep{2013A&A...558A..33A}. The use of aperture photometry is appropriate as the images of the nova are taken with very short exposures and there are no closer sources in the vicinity of the nova.

We estimated the magnitudes for the reference star located in the field of view of the nova using the mean value of five nights of observations. Our photometry for individual nights was computed using the star instrumental magnitude and the $BVRI$ zero points derived from observations of photometric standards, as described in \citet{2018ApJ...863...90W}. We derived the optical zero points using the Landolt standards in the PG1657+078 field \citep{1992AJ....104..340L}. The infrared zero points are calibrated using 2MASS $JHK$ data \citep{2MASS_PSC_2006}. The zero points are extinction and color-term corrected based on the data provided on the CTIO's calibration pages \footnote{\url{http://www.ctio.noao.edu/noao/content/13-m-smarts-photometric-calibrations-bvri}} \footnote{\url{http://www.ctio.noao.edu/noao/content/photometric-zero-points-color-terms}}.

Due to the brightness of the nova ($V_{\rm{max}}$ = 5.84\,mag), only one star is suitable to be used as a reference star in the optical, as it is bright enough to be visible alongside the nova when the object is at its maximum and dim enough to not be saturated when we increased the exposure time during the later part of the observation. For these reasons, we did not use more reference stars for our optical photometry than the one labeled. We checked the reference star for non-variability through the data from the Gaia catalog \citep{refId01,GaiaDR2} and found that the star has a nominal magnitude error and insignificant astrometric noise. There is hence strong evidence that the reference star is not a variable and is appropriate for our photometry calibration purposes.

\begin{figure*}
\plotone{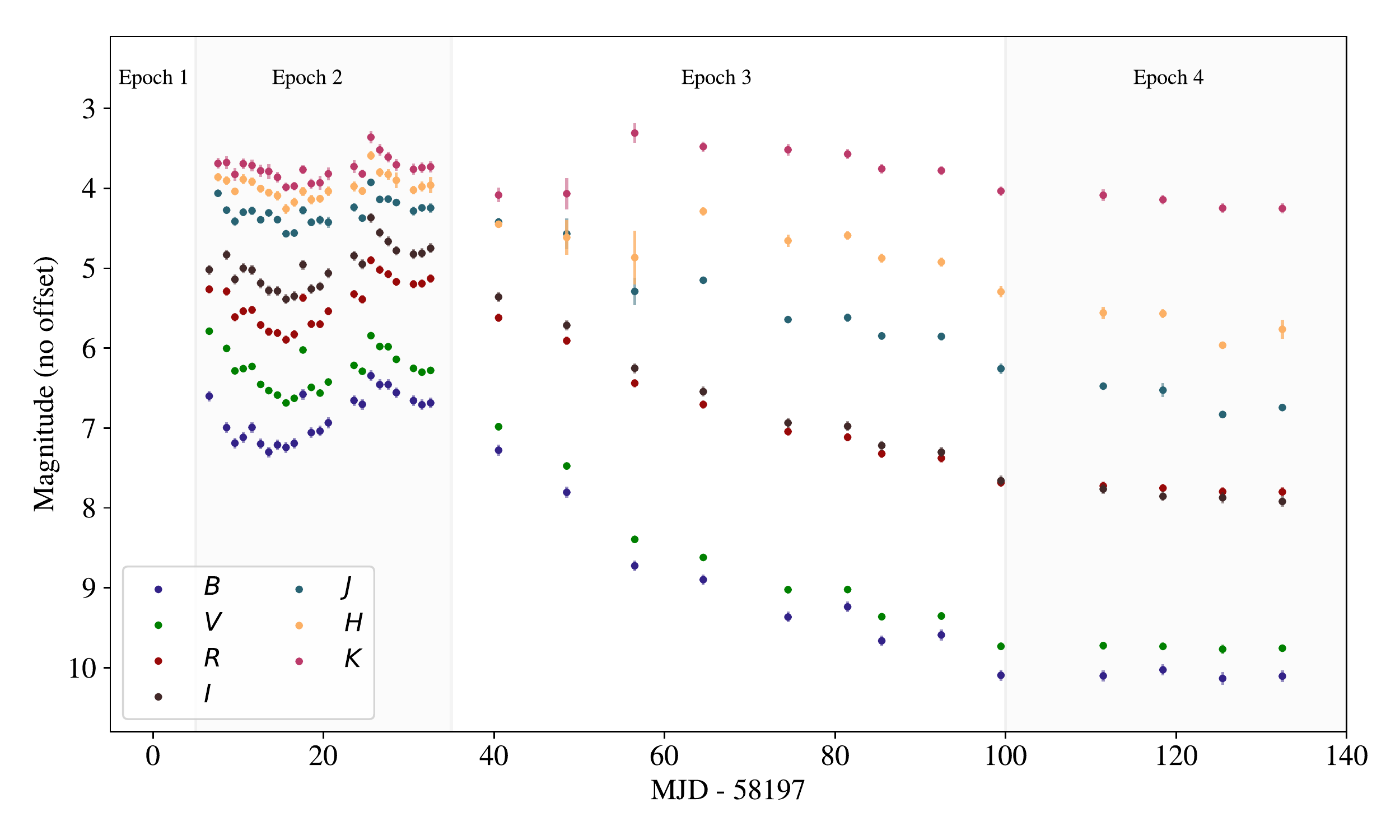}
\caption{Light curves of V906 Car in the optical and near-infrared in their constituent epochs. No offset is applied.\label{fig:fullc}}
\end{figure*}

The near-IR data are reduced using the astronomy software \texttt{Cyanogen Imaging MaxIm DL}'s photometric tool, which has the in-built capability to identify, track, and photometer objects across images using aperture photometry. We used the same star in the near-IR $JHK$ filter magnitudes as we did for the optical images. We present the $BVRIJHK$ light curves in Figure \ref{fig:fullc}. 

\begin{deluxetable*}{ccccccc}
\tablecaption{Optical Photometry Sequence near V906 Car \label{tab:opseq}}
\tablewidth{0pt}
\tablehead{
\colhead{Star ID} &
\colhead{$\alpha$ (J2000)} &
\colhead{$\delta$ (J2000)} & 
\colhead{$V$} & 
\colhead{$B-V$} &
\colhead{$V-R$} &
\colhead{$V-I$}
}
\startdata
  Nova & 10\textsuperscript{h}36\textsuperscript{m}15\textsuperscript{s}.239 &  $-59^{\circ}$35'52 326" & ... & ... & ... &... \\
  1 & 10\textsuperscript{h}36\textsuperscript{m}22\textsuperscript{s}.371 & $-59^{\circ}$36'31 636" &9.185 (031)& 1.782 (032) & 1.268 (048)&2.267 (055)\\
\enddata
\end{deluxetable*}

\begin{deluxetable*}{ccccc}
\tablecaption{Infrared Photometry Sequence near V906 Car \label{tab:irseq}}
\tablewidth{0pt}
\tablehead{
\colhead{Star ID} &
\colhead{$J$} & 
\colhead{$H$} & 
\colhead{$K$} 
}
\startdata
  1 &5.423 (004)&4.668 (0135)& 4.304 (011)\\
\enddata
\end{deluxetable*}

\begin{figure}
\includegraphics[width=0.48\textwidth]{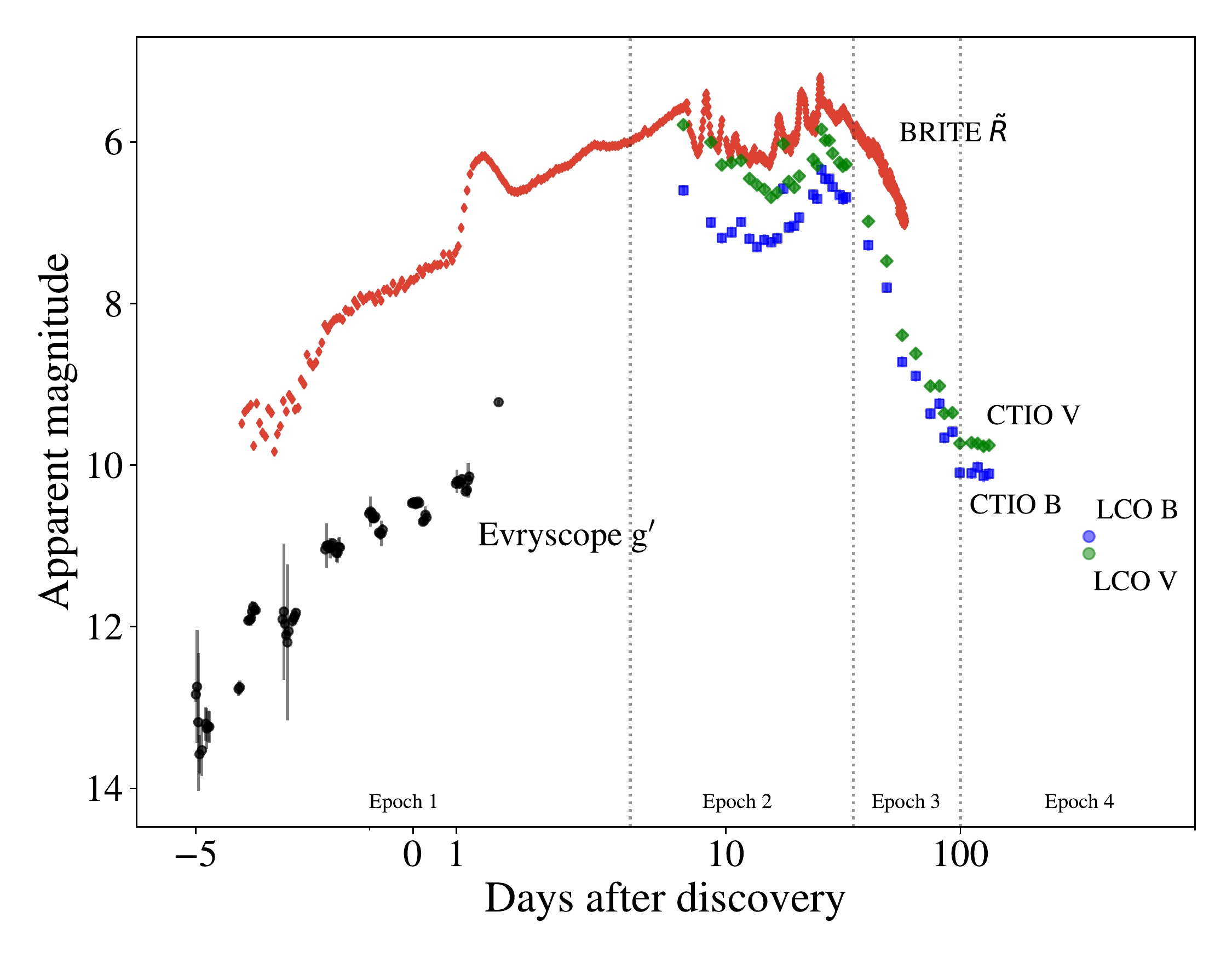}
\caption{Pre-discovery early time data from Evryscope. BRITE $\tilde{R}$ data is superimposed on CTIO $B$ and $V$ photometry. Late-time LCO $B$ and $V$ data are also added.} No offsets are applied. \label{fig:evryscope} 
\end{figure}

\begin{deluxetable*}{cccccccccc}
\tablecaption{Optical and Near-IR Photometric Cadence of V906 Car. All magnitudes are reported in the Vega system. \label{tab:cadence}}
\tablecolumns{9}
\tablehead{
\colhead{Julian Date} &
\colhead{} & 
\colhead{} &
\colhead{} & 
\colhead{} & 
\colhead{} &
\colhead{} &
\colhead{} & 
\colhead{} \\
\colhead{(2458000)} &
\colhead{$B$} & 
\colhead{$V$} &
\colhead{$R$} & 
\colhead{$I$} & 
\colhead{$J$} &
\colhead{$H$} & 
\colhead{$K$} 
}
\startdata
203.600 & 6.600(134) & 5.788(091) & 5.265(106) & 5.020(125) & .... & .... & ....\\
204.623 & .... & .... & .... & .... & 4.497(04) & 4.290(051) & 3.910(067)\\
205.636 & 6.995(134) & 6.004(091) & 5.290(106) & 4.832(125) & 4.708(045) & 4.333(050) & 3.898(082)\\
206.612 & 7.189(134) & 6.285(091) & 5.612(106) & 5.140(125) & 4.849(058) & 4.468(046) & 4.047(083)\\
207.587 & 7.117(134) & 6.257(091) & 5.538(106) & 5.000(125) & 4.734(048) & 4.317(064) & 3.913(067)\\
208.619 & 6.992(134) & 6.230(091) & 5.522(106) & 5.023(125) & 4.715(053) & 4.347(050) & 3.935(076)\\
209.628 & 7.199(134) & 6.454(091) & 5.710(106) & 5.187(125) & 4.829(044) & 4.433(050) & 4.001(075)\\
210.581 & 7.301(134) & 6.532(091) & 5.793(106) & 5.279(125) & 4.742(044) & 4.483(051) & 4.009(095)\\
211.602 & 7.212(134) & 6.587(091) & 5.811(106) & 5.285(125) & 4.827(046) & 4.523(060) & 4.083(066)\\
.... & .... & .... & .... & .... & .... & .... & .... & .... \\
\enddata
\tablecomments{Julian dates refer to the median Julian dates for all the images taken in a night. This table is truncated. The complete dataset is fully reproduced in the digital form.}
\end{deluxetable*}

\subsection{CTIO 1.5m CHIRON} \label{subsec:spectra}

We observed V906 Car with the CTIO 1.5-meter telescope and CHIRON spectrograph
 \citep{2012SPIE.8446E..0BS,2013PASP..125.1336T} over $70$ days in the ``slicer'' mode whose spectral resolution is up to $R=$80,000. Upon publication, the data will be made publicly available via the online repository \texttt{WiseREP} \citep{2012PASP..124..668Y}.

\subsection{Evryscope} \label{subsec:earlytime}

We obtained pre-discovery $g'$-band photometry of the nova from Evryscope, spanning the time between $T_{D}-5$ and $T_{D}+2$. 
Evryscope  is  an  array  of  telescopes  capable of observing the entire visible sky down to an airmass of 2 (8,150 square degrees), together forming a gigapixel-scale image every 2 minutes \citep{2015PASP..127..234L, 2019PASP..131g5001R}. 
Data is typically taken at a 2 minute cadence, occasionally twice-over if the object appeared in two cameras simultaneously. 
The images were co-added to a 30 minute cadence for better consistency in the PSF. 
The Evryscope $g'$ pre-discovery light curve is shown in relation to our $B$ and $V$ light curves in Figure~\ref{fig:evryscope}.

 Evryscope data is reduced in real time by a custom analysis pipeline described in \citet{2019PASP..131g5001R}. 
 After background modelling and subtraction, differential photometry is performed with forced apertures at known source positions in a reference catalog \citep{2019PASP..131g5001R}.
 The \texttt{SysREM} detrending algorithm is applied with two iterations to remove any remaining systematics \citep{2005MNRAS.356.1466T}. 

\subsection{LCO 1.0m} \label{subsec:lco}
We obtained a late-time optical photometry of the nova approximately a year after discovery (MJD 58550.2) using the LCO 1.0\,m telescope under the program NOAO2019A-011 (PI: N. Blagorodnova). The data were reduced using the LCO \texttt{BANZAI} pipeline \citep{BANZAI_curtis_mccully_2018_1257560}. Aperture photometry on the target was performed using \texttt{Python photutils} routines. The zeropoint for $g$ and $r$ bands were calibrated using isolated field stars that were available in the SkyMapper DR1.1 catalogue \citep{SkyMapper_Wolf2018PASA}. The zeropoints for $B$ and $V$ bands were derived using field stars present in the UCAC4 catalogue \citep{UCAC4_2012yCat}. The rest of filters was calibrated using as a reference the star from Table \ref{tab:opseq}.

\section{Extinction estimate} \label{sec:high-res_spec}

Using the Milky Way extinction law from \cite{Fitzpatrick1999PASP} and the extinction parameters from \cite{Schlafly2011ApJ}, we obtain an $A_V= 3.598$ for the position of the nova. However, this shall be only considered as an upper limit, as this is the total line of sight extinction in our own Galaxy; the nova, located approximately at the distance of the Carina Nebula, has likely less extinction. In order to obtain a better estimate for the foreground extinction to our target, we used the strength of Na\textsc{i} lines from high resolution spectroscopy of the nova as a proxy for the extinction.

We used the high resolution sectrum taken with the CTIO 1.5-meter telescope to measure the absorption in the wavelengths of Na\textsc{i} D\textsubscript{1} and Na\textsc{i} D\textsubscript{2} using the \texttt{IRAF} spectral analysis routines. The absorption was assumed to arise from foreground interstellar medium. After measuring the equivalent widths, we computed the average values for both the D\textsubscript{1} and D\textsubscript{2} lines. Our sample of \io{Na}{i} equivalent width measurements was converted into $E(B-V)$ using the empirical relations found in \cite{2012MNRAS.426.1465P}. We also calculated $A_V$ using the assumed ratio for total to selective extinction $R_V$ of 3.1 \citep{1989ApJ...345..245C}. The result of our analysis suggests a foreground extinction toward V906 Car of $A_V=1.116_{-0.385}^{+0.538}$ mag, which is likely unaffected by the peculiar reddening law inside the nebula \citep[e.g. $R_V=4.8$;][]{Smith1987MNRAS}.
Our value is in good agreement with the foreground extinction values towards Carina derived in recent studies \citep{Hur2012AJ}, which estimated a relatively low reddening $E(B-V) \sim 0.36 \pm 0.04$, and extinction of $A_V \sim 1.1$\,mag. These values also agree with the spectroscopic analysis done by \citet{2020NatAs.tmp...79A}, which used combined Na\textsc{i} D\textsubscript{1} and D\textsubscript{2} interstellar absorption doublet method and diffuse interstellar bands (DIBs) to derive a foreground extinction towards the nova of $A_V = 1.11 \pm 0.05$ mag.

\section{Photometric evolution of V906 Car} \label{sec:photevol}

\subsection{Evolution of Optical and Near-IR Emission} \label{subsec:evolution}

The nova conforms to the general morphology of a nova light curve as described by \citet{2008clno.book.....B}. The nova is observed to have an initial rise from $t_{D-5}$ to $t_{D+7}$, a pre-max halt at $t_{D+6}$, followed by a final maximum on $t_{D+26}$, an early decline from $t_{D+26}$ to $t_{D+100}$, and finally a slow down in the decline after $t_{D_{>100}}$. 

The photometric evolution of V906 Car can be similarly grouped into four epochs: (1) the rising part from $t_{D-5}$ to $t_{D+2}$; (2) the evolution at peak, lasting from $t_{D+5}$ to $t_{D+35}$; (3) the decay phase from $t_{D+35}$ to $t_{D+100}$; and (4) the late plateau phase, which begins at $t_{D+100}$.

In the first epoch, the Evryscope data (Figure~\ref{fig:evryscope}) shows that the nova rose from $g'=13.58\pm0.23$ to $g'=9.22\pm0.03$ magnitudes in 6 days. The light curve shows minor bumps of fractions of a magnitude along a power law rise. The non-smooth uniform rise seems to qualitatively agree with the pre-peak high cadence data obtained by BRITE. 

In the second epoch, the light curve begins with an immediate decline from our first day of observations. The decline lasts for 10 days in all filters, falling most significantly in the $B$ and $V$ with a $\sim$1\,mag decline. The dip, however, was not significant in the $JHK$ filters. After this initial decline, the nova re-brightened to its peak in all $BVRIJHK$ filters, with $V_{\rm{max}} = 5.84 \pm 0.09$ mag. Another interesting feature is that the early period is punctuated with sudden outbursts, with the most pronounced flare on $t_{+15}$, where the nova quickly brightened for about a day before falling back to a slower rise in all filters. The flare was most significant in the optical filters, brightening by 0.6 magnitudes in the $V$-band. 

In the third epoch, the brightness of the nova starts to decline at $t_{+35}$. The decline lasts until $t_{100}$ for all bandpass filters except for $H$ and $K$ bands, which begin to brighten at around $t_{50}$, before reaching a peak and declining thereafter at $t_{60}$. Notably, the $K$ magnitude reached a maximum brighter than the initial maximum in the early epoch. This change is probably unrelated to dust formation, as we do not see a significant rise in the other near-IR bands. A more likely explanation is an increase in the flux in the He\,\textsc{i} 2.05 $\mu$m line emission \citep{1996ApJ...456..717S}. The larger excess in $K$ is consistent with expectations that the thermal emission will peak in or beyond the $K$ band.

In the fourth and final epoch, the brightness of the nova remains largely flat in the $BVRI$ filters until the end of our observations. The brightness of the nova continue to decline in the $JHK$ bands, with $J$ and $H$ declining at a similar rate while $K$ declines at a much slower rate.

\subsection{Colour evolution} \label{subsec:colors}

\begin{figure}
\includegraphics[width=0.45\textwidth]{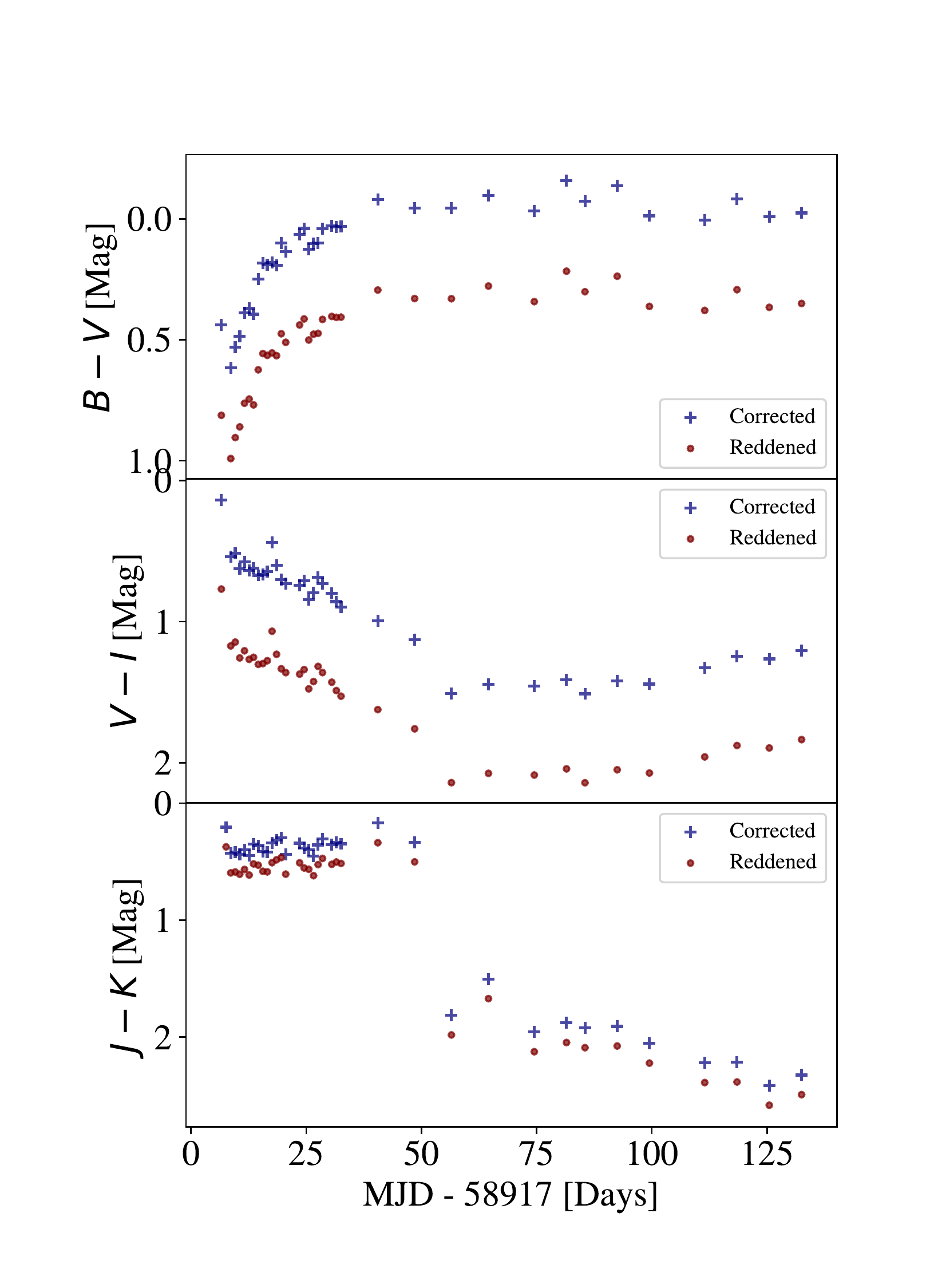}
\caption{Color evolution $B-V$, $V-I$, and $J-K$ of V906 Car. Extinction corrections are based on the extinction value of $A_V = 1.11$. \label{fig:bv}}
\end{figure}

We present the evolution of the nova optical and infrared colors $B-V$, $V-I$, and $J-K$ in Figure \ref{fig:bv}. In the $B-V$ curve, the nova becomes bluer from day 10 to day 45 after outburst, and then levels off to a nearly constant value of $B-V=0.24$ for the remaining observations. The bluer colors are accompanied by a steady reddening in $V-I$, which changes from $V-I=1.3$ to $V-I=2.3$ during the same period. The infrared color index $J-K$ shows a quick change near day 50, and shifts from steady values of $J-K=0.7$ observed prior to day 50 to values above $J-K=2$ after day 50. At later times, while the optical $B-V$ colour remains stable and the $V-I$ colour gets bluer, the IR bands slowly evolve towards redder colours.

We interpret these changes in optical and infrared colour in terms of the thinning of the photospheric optical depth, which would result in an unobstructed view of the hotter accretion disk and the white dwarf, having bluer $B-V$ colors. The later production of dust would cause a gradual increase in $V-I$ colors from increasing infrared emission from heated dust. The sudden increase in $J-K$ colors at day 45 could be explained by the formation of this dust envelope within the nova, which over time causes an increasing infrared excess from the source.

\section{Analysis} \label{sec:analysis}

The progenitor of V906 Car lies at a closer angular separation from the Carina Nebula, suggesting a possible association. The distance to the nebula was estimated  to be of $d=2.3 \pm 0.5$\,kpc using the proper motion of the Homunculus Nebula \citep{Smith2006ApJ}. The distance for the progenitor star as provided in the \textit{Gaia} DR2 distance catalogue \citep{GaiaDR2,Bailer-Jones2018AJ} is $d=3.35_{-1.60}^{+2.81}$\,kpc. Although the errors are large due to the faint magnitude of the star, this value seems to be consistent with being associated with the nebula. In fact, our estimate for the extinction is significantly lower than the total value for the line of sight. Because the nebula is likely to contribute significantly to the overall reddening value, in our analysis we will assume that the nova is most likely located in front of or inside the nebula, and will adopt the nebula's distance for the progenitor.

\begin{table}
\begin{minipage}{1.\linewidth}
\begin{small}
\caption{Photometry of the progenitor for V906 Car. All magnitudes are reported in the Vega system.}
\centering
\begin{tabular}{cccc}
\hline
Survey & Band  & Magnitude & Reference    \\ 
    &  		& (mag) &     \\ \hline
 VPHAS+ 	& $u$ & 19.46$\pm$0.05& [1] \\
 VPHAS+ 	& $g$ & 20.13$\pm$0.03& [1] \\
 VPHAS+ 	& \halpha & 19.09$\pm$0.05& [1] \\
 VPHAS+ 	& $r$ & 19.62$\pm$0.03& [1] \\
 VPHAS+ 	& $i$ & 19.18$\pm$0.04& [1] \\
 VISTA 		& $J$ & 18.58$\pm$0.06& [2] \\
 VISTA 		& $H$ & 18.41$\pm$0.10& [2] \\
\hline
\end{tabular}
\begin{tablenotes}
\item[\textdagger]$^a$  References: [1] \cite{Drew2014MNRAS_VPHAS,Drew2016yCat}, [2] \cite{Preibisch2014AA_VISTA}.
\end{tablenotes}
\label{tab:progmag}
\end{small}
\end{minipage}
\end{table}

\subsection{Progenitor modeling} \label{sec:progenitor}

The location of the nova was covered by the VST Photometric H$\alpha$ Survey of the Southern Galactic Plane and Bulge \cite[VPHAS+;][]{Drew2014MNRAS_VPHAS}, performed between February 2012 and April 2013 (7$-$8 years before the nova outburst) and the Visible and Infrared Survey Telescope for Astronomy \citep[VISTA;][]{Preibisch2014AA_VISTA}, which observed the area in March 2012. 

We used the online portal \texttt{Vizier}\footnote{\url{http://vizier.u-strasbg.fr/viz-bin/VizieR}} to retrieve the archival magnitudes of the progenitor system, which are shown in Table \ref{tab:progmag}.

\begin{figure}
\hspace{-0.5cm}
\includegraphics[width=1.05\linewidth]{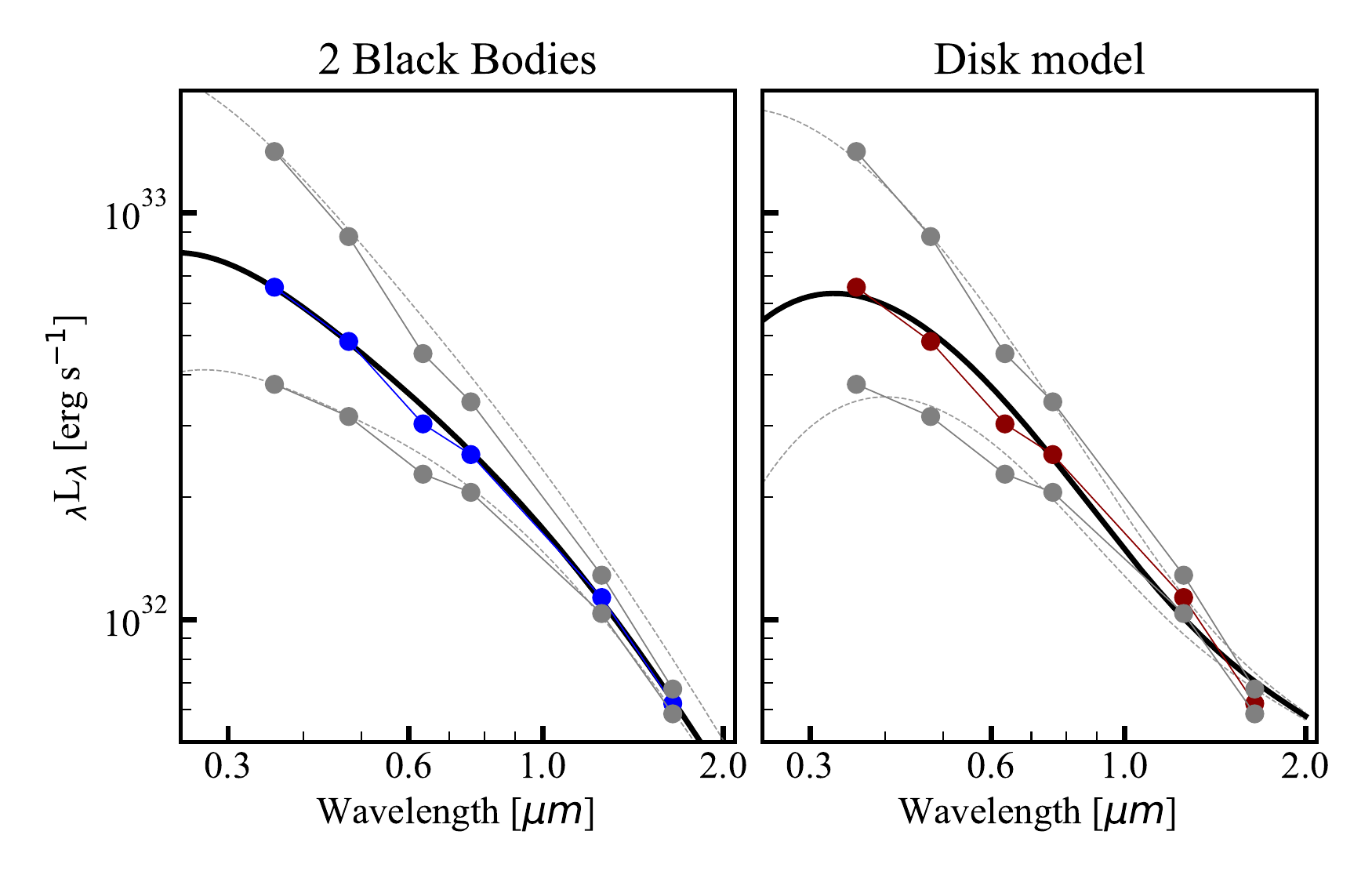}
\caption{Best fit to the progenitor SED with a two black-body model (left) and a steady accretion disk model (right). The filled points correspond to the archival flux for the progenitor in $ugriJH$ bands. The black line corresponds to the best fit model for $A_V=1.116$. The gray markers show the SED with extinction correction values of $A_V$=0.73 and $A_V$=1.65\,mag. The dashed gray lines correspond to the best fit models for those SEDs. \label{fig:sed_progenitor}}
\end{figure}

The modeling of the progenitor was performed using a custom developed code \texttt{BBFit}\footnote{\url{https://github.com/nblago/utils}}, based on Markov-Chain Monte Carlo (MCMC) \texttt{Python} implementation in the package \texttt{emcee} \citep{Foreman-Mackey2013PASP}. After correcting for the foreground extinction (see \S \ref{sec:high-res_spec}), we used the transmission curves for each filter ---available through the Spanish VO Filter Profile Service \citep{SVO_2012}--- to obtain the monochromatic flux for each band. Initially, we modeled the emission with two black-body components. Due to an excess of flux in the \halpha band, likely due to an existing emission line, we left this band out of the analysis. The best-fit model for different values of extinction is depicted in Figure \ref{fig:sed_progenitor} and the posterior parameters are summarized in Table \ref{tab:model_fit}.

Our results show that the hot component has a temperature of $\sim 1.5 \times 10^4$\,K and a luminosity of $\sim 3 \times 10^{-1}$\,\Lsun. Its estimated radius of $\sim 50 \times 10^8$\,cm is larger than the radius of a single WD, which ranges $0.5 \times 10^8 \leq R_{WD} \leq 14 \times 10^8$\,cm. These values are in agreement with a bright spot or an accretion disk, rather than the WD itself.

The cold component has a luminosity of $(1.6-4) \times 10^{-2}$\,\Lsun, consistent with a cold dwarf classification. However, its estimated temperature is higher than the one expected for a main sequence star within this luminosity range. Most likely, this is due to a reflection effect, when in a tidally locked system the radiation from the WD heats one of the sides of the cold companion. 

Given the donor is a main sequence star, we use the mass-to-radius and the mass-to-luminosity relations to estimate its mass. For our canonical extinction value, the  $R/R_{\odot}=(M/M_{\odot})^{0.8}$ relation suggests a donor with $\simeq$0.25\,\Msun. When using the mass-luminosity relation $L/L_{\odot} \simeq 0.23 (M/M_{\odot})^{2.3}$ we derive an upper limit for the companion mass of $\simeq$0.43\,\Msun.

\begin{table*}
\begin{minipage}{1.\linewidth}
\begin{small}
\caption{Best-fit parameters for the progenitor SED modelling. }
\setlength{\tabcolsep}{3pt}
\centering
\begin{tabular}{ c | cccccc | cccc }
& \multicolumn{6}{c|}{2 Black Bodies}  & \multicolumn{4}{c}{Disk Model} \\ \hline
Av & T$_{BB1}$  & R$_{BB1}$ & L$_{BB1}$  & T$_{BB2}$  & R$_{BB2}$ & L$_{BB2}$  & $M_*$ & $R_*$ & log  & $R_{\rm{out}}$ \\ 
(mag) & (K) & (10$^8$\,cm) & (10$^{-1}$ $L_{\odot}$) & (K) & (10$^8$\,cm) & (10$^{-2}$ $L_{\odot}$)& ($M_{\odot}$) & (10$^8$\,cm) &  [$\dot{M}$ /($M_{\odot} \rm{yr}^{-1}$)] &  ($R_{\odot}$)  \\ \hline
0.731 & 13936$^{+2188}_{-1395}$ & 45$^{+9}_{-9}$ & 1.4$^{+2.3}_{-0.8}$ & 5067$^{+319}_{-351}$ & 179$^{+9}_{-7}$ & 4.0$^{+1.6}_{-1.2}$  &   0.72$_{-0.23}^{+0.23}$ & 27.2$_{-0.5}^{+0.5}$ & $-9.7_{-0.1}^{+0.2}$ & 20.7$_{-0.6}^{+0.6}$ \\           
1.116 & 15073$^{+2124}_{-1450}$ & 54$^{+9}_{-9}$ & 2.8$^{+3.8}_{-1.6}$ &  4951$^{+497}_{-527}$ & 172$^{+14}_{-11}$ & 3.3$^{+2.4}_{-1.5}$ & 0.71$_{-0.19}^{+0.23}$ & 21.2$_{-0.5}^{+0.5}$ & $-9.7_{-0.1}^{+0.2}$ & 22.1$_{-0.6}^{+0.7}$  \\          
1.654 & 18481$^{+1502}_{-1040}$ & 61$^{+4}_{-5}$ & 8.4$^{+4.7}_{-2.9}$ & 3904$^{+616}_{-551}$ & 192$^{+38}_{-25}$ & 1.6$^{+2.5}_{-0.9}$  & 0.73$_{-0.22}^{+0.22}$ & $13.1_{-0.4}^{+0.4}$ & $-9.7_{-0.1}^{+0.2}$ &24.9$_{-0.7}^{+0.7}$ \\  \hline
\end{tabular}
\label{tab:model_fit}
\end{small}
\end{minipage}
\end{table*}

Provided the nova progenitor was an accreting WD system, we the undertake a second analysis of the observed SED using the formalism of a viscous accretion disk model developed by \cite{ShakuraSunyaev1973AA} and \cite{Lynden-BellPringle1974MNRAS}. In this approach, we assume that the emission is dominated by the accretion disk and therefore the integrated SED of the source is composed by several concentric optically thick annuli, each one radiating at a different temperature. Assuming that the accretion rate through the disk is steady, the energy dissipated within each annulus is independent of the viscosity. For a specific implementation, we adopt the model described in \cite{Frank2002apa..book}, where the temperature of each annulus at radius $R$ corresponds to

\begin{equation}
    T(R) = \left\{ \frac{3 G M \dot{M}}{8 \pi R^3 \sigma} \left[ 1 - \left( \frac{R_*}{R} \right) ^{1/2}   \right]  \right\} ^{1/4}
\end{equation}
where $M_*$ is the mass of the star, $R_*$ the stellar radius, $\dot{M}$ the the accretion rate and $\sigma$ and $G$ are the Stefan-Boltzmann and gravitational constants. Following the approach of \cite{Kenyon1988ApJ}, which used the steady disk approach to study FU Orionis, we assume that the disk emits at its maximum temperature for any radius $R_{*} \leq  R \leq 1.5R_{*}$. 

The spectrum for each annul is approximated as a black-body at effective temperature $I_{\nu} = B_{\nu} [T(R)]$, which is integrated up to the outer radius of the disk $R_{\rm{out}}$. The binary inclination angle is designed by $i$, which we set to a fix value of $i=45^{\circ}$. The emission from each ring of thickness $dR$ is scaled by the solid angle as seen from distance $D$, $2 \pi \,R\,$d$R$\,cos\,$i / D^2$.  This provides the total monocromatic flux to be

\begin{equation}
    F_{\nu} = \frac{4 \pi h \, \rm{cos} \,i \,\nu^{3}}{c^2 D^2} \int_{R_*}^{R_{\rm{out}}} \frac{R dR}{e^{h \nu / k T(R)} - 1}
\end{equation}

This accretion disk model has four different parameters, for which we assume uniform priors: $0 \leq M_* \leq 1.44$\,\Msun, $0 \leq R_* \leq 10$\,\Rsun, $-10 \leq \rm{log} [\dot{M}/(M_{\odot} yr^{-1})] \leq -5$ and $R_* \leq R_{\rm{out}} \leq 50$\,\Rsun.

Adopting different foreground extinction values for the system, the best fit parameters of the disk model are shown in Table \ref{tab:model_fit}. For our best estimate of the extinction $A_V=1.116$, the computed stellar radius is still a factor of two larger than the one expected for white dwarfs of the given mass. However, deviations are not unexpected, due to the highly simplified disk model used in our analysis and ignoring the contribution of the cold donor star. As well, assuming a larger value for the extinction $A_V$ also allows us to obtain a more consistent radius, while all the other parameters are not significantly altered.

The best-fit SEDs for both models are shown in Figure \ref{fig:sed_progenitor}. The two black-body model seems to already work reasonably well in predicting both the blue flux from the disk and the colder contribution from the companion. The composite SED of the accretion disk seems to over-predict the flux at longer wavelengths, which translates in a larger $R_{\rm{out}}$ than the one physically expected for such system. The progenitor mass that we obtain with the accretion disk modeling does not show a strong dependence on the extinction. This estimate is also in agreement with the WD mass derived from our analysis of the decay of the light curve, as further explained in \S \ref{subsec:wdmass}.

\subsection{Evolution of the V906 Car eruption}

\begin{figure*}
\plotone{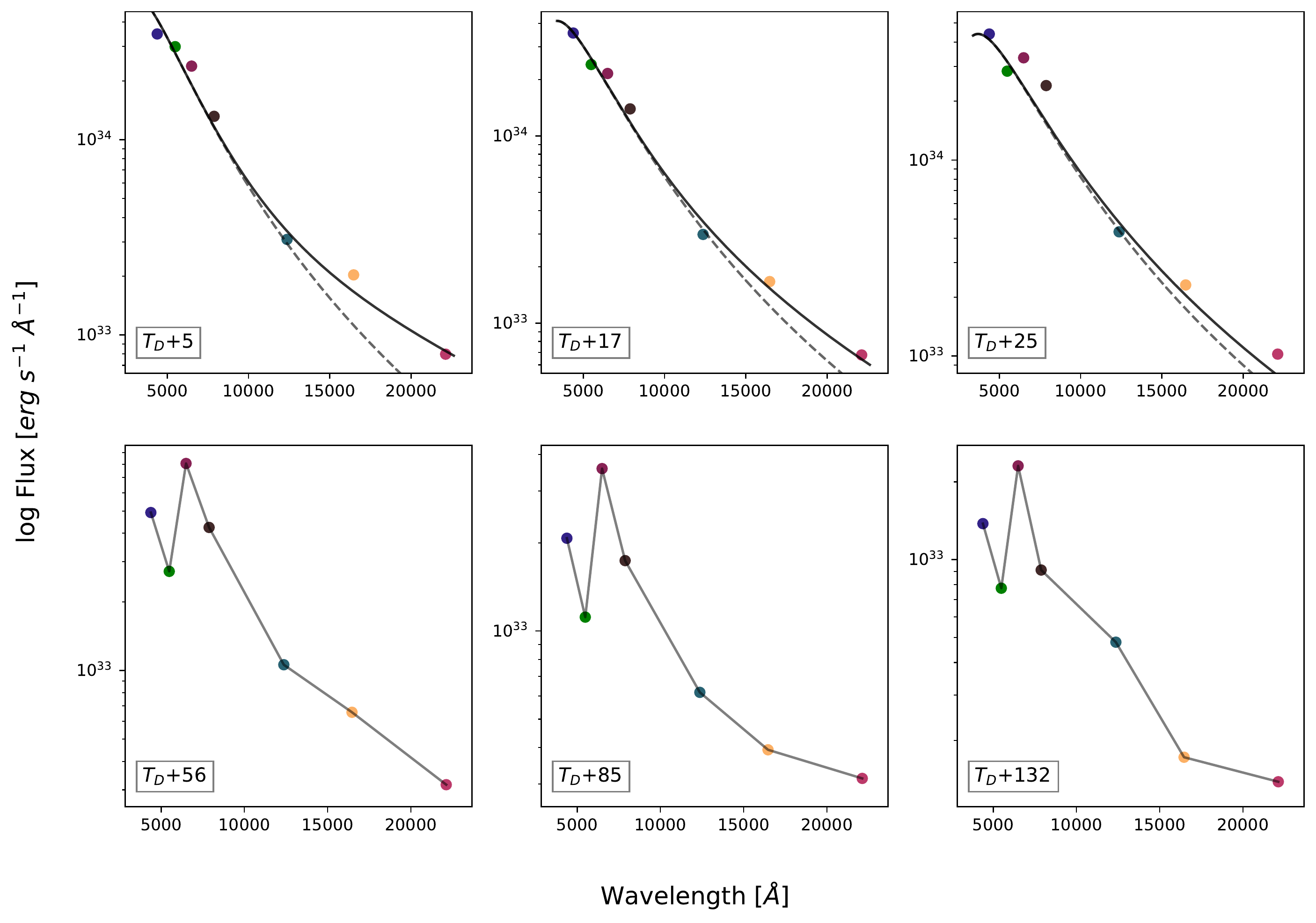}
\caption{Evolution of the Spectral Energy Distribution of the nova. The fluxes at different epochs after discovery are shown for $BVRIJHK$ bands. The dotted lines represent single black body fit, while the solid lines represent best fit for two black body model. After $t_{50}$ the SED is dominated by strong emission lines, which cause the SED to deviate from a black-body.  \label{fig:sed_evolution}}
\end{figure*}

\begin{figure}
\plotone{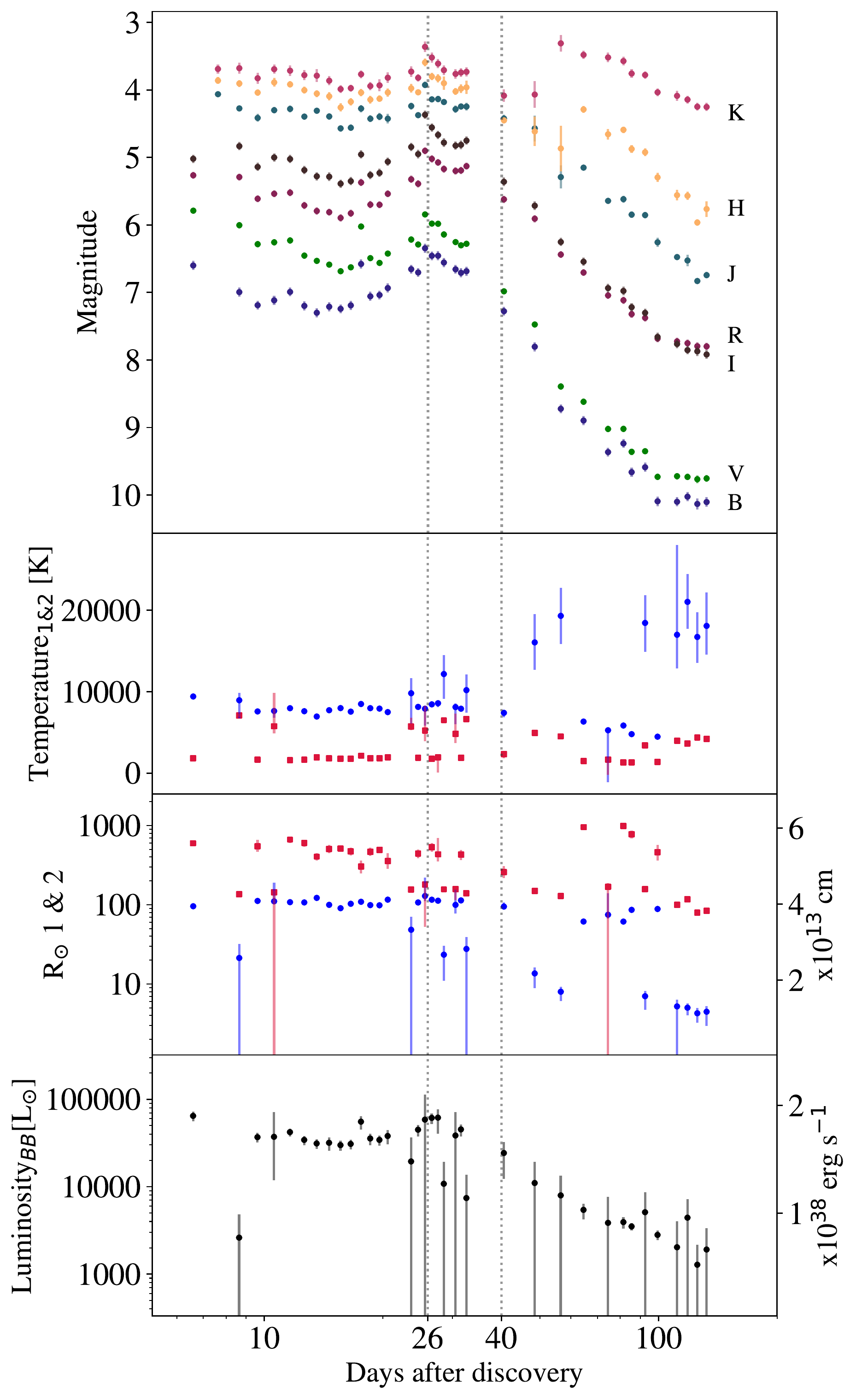}
\caption{Temperature, radii and luminosity of two black-body fit modelling over time. The hotter component is shown with blue markers and the colder components with green. The dotted line indicates the transition of the different phases in the nova SED evolution. 
\label{fig:trl_compare}}
\end{figure}

Using the same assumptions for distance and extinction as in Section \ref{sec:progenitor}, we used \texttt{BBFit} to model the nova SED at different phases, shown in Figure \ref{fig:sed_evolution}. Since, as mentioned earlier, soft X-ray emission was not detected from the nova, we worked only with optical and near-IR data and did not include X-ray or UV data in our modelling. Provided the existence of an IR excess component in nova eruptions, we chose to use two black-bodies to model the emission.

The best fit measurements from our analysis are shown in Figure \ref{fig:trl_compare}, along with the multi-band photometry. We divide our analysis into three different phases based on the evolution of the SED. In the first phase, from $t_D$ to $t_{D+25}$, the nova is dominated by the optically thick wind, which agrees fairly well with a black-body description. 

The temperature and radii derived using this simple model remain fairly consistent. The average temperature of the hotter black-body is $\sim$9540\,K and a its photospheric radius is $\sim$70R\textsubscript{\(\odot\)}, while the colder is at 2440\,K throughout the first 25 days (epoch2), with a radius of $\sim 500$R\textsubscript{\(\odot\)}. The luminosity of the system is also fairly consistent in these 25 days, with a median of 3.6$\times10^4$\,\Lsun (or average of 3.96$\times10^4$\,\Lsun). Since the peak of the emission is at shorter wavelengths, without UV data, it is difficult to exactly constrain the SED. Therefore, our estimate of the nova's bolometric luminosity should be interpreted as a lower limit without including the emission at shorter wavelengths.

In the second phase of the SED evolution, from $t_{D+25}$ to $t_{D+40}$, the source starts to develop an optically thin wind with strong emission lines, specially in the $R$ and $I$ band region (see Figure \ref{fig:sed_progenitor}), which makes the black-body fits less reliable. In our attempt to minimize the effect of the lines, we decide to exclude these bands from our fit. However, the larger scatter and increased error bars on the best fit parameters indicate the these values should be interpreted with caution.

In the third and fourth phases, from $t_{D+40}$ onward, the non black-body like appearance of the nova SED makes the interpretation challenging, as shown in Figure \ref{fig:sed_evolution}. The emission is not well approximated by a simple black body model. At this epoch, the hotter component increases in temperature to $\sim2 \times 10^4$\,K, as the photosphere recedes closer to the WD. The bolometric luminosity shows a smooth decline. While the fits qualitatively agree with the nova expected behaviour, a more careful future analysis based on spectroscopic modelling is required.

\section{Discussion} \label{sec:discussion}

\subsection{Nova Classification} \label{subsec:classification}

From the evolution of V906 Car in the visual band, we derive the times for the nova to fade by 2 and 3 magnitudes, $t_{2,V} = 26.2$ days and $t_{3,V} = 33.0$ days. The decline rate (speed) makes V906 Car a ``fast nova'' according to the \citet{1957gano.book.....G}  classification. Categorising CNe in speed classes is, however, for most parts only useful taxonomically. A nova's astrophysical significance can be better understood in relation to the classification method by \citet{Strope2010AJ}, which categorizes CNe according to their light curve morphology.

V906 Car exhibits characteristics of various nova classification in terms of light curve shape. An early photometric analysis by \citet{2018ATel11677....1D} determined V906 Car to be a J class nova, due to the `jittering' above the base level of the nova light curve shape. \citet{2018ATel11677....1D} also considered the potential that V906 Car could be classified as a O class nova, but commented that the oscillations in a O class nova are supposed to be quasi-periodic, and the periodicity of the jitters in V906 Car appeared to be random.

\begin{figure}
\includegraphics[width=0.48\textwidth]{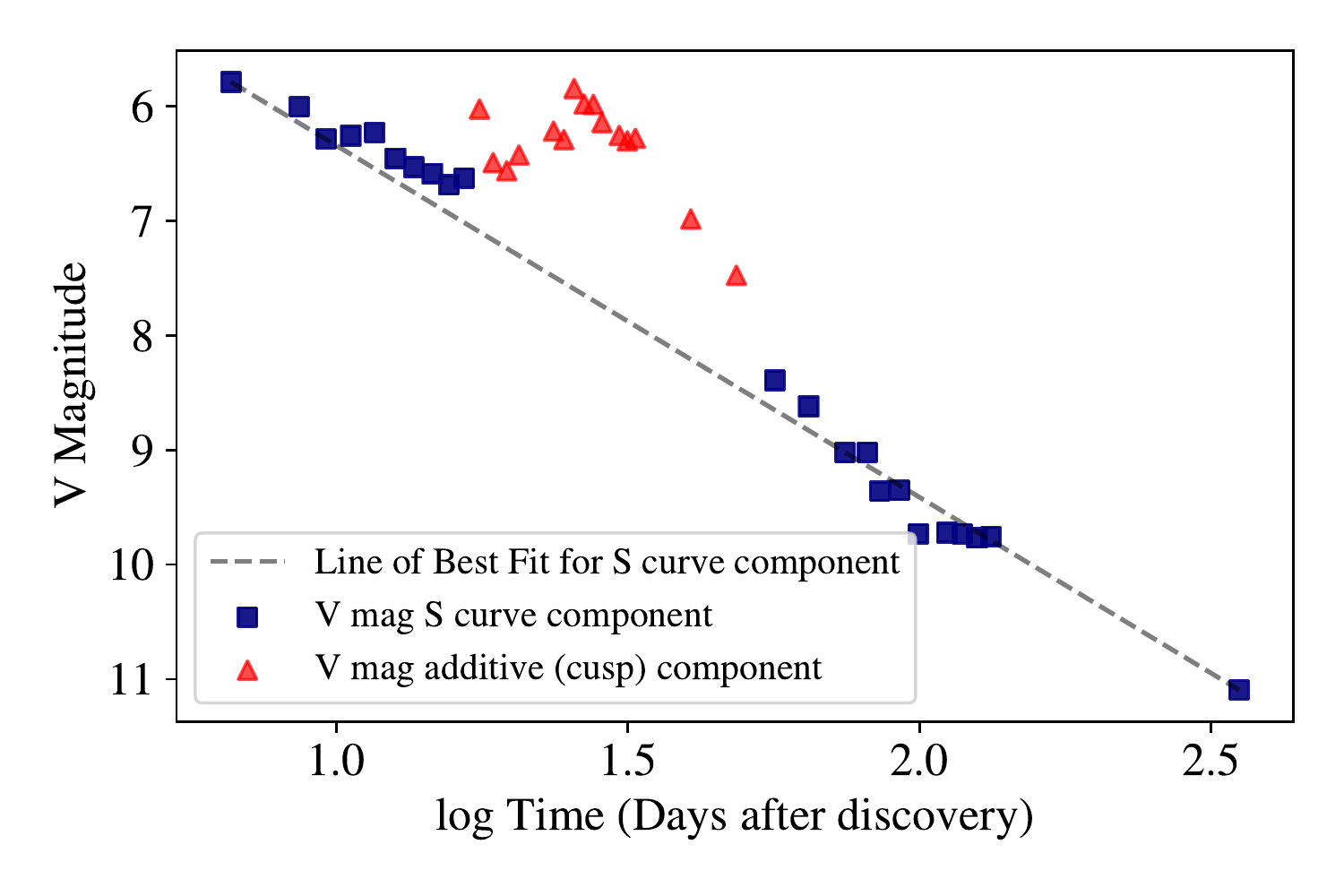}
\caption{Underlying smooth nova characteristics fitted with trend line, emphasizing additive component. \label{fig:cusp_nova}}
\end{figure}

The analysis by \citet{2018ATel11677....1D} is based on a limited profile of the nova, with less than 3 months of data. Using our combined CTIO and LCO data spanning for about a year, we consider V906 Car to most closely resemble a C class nova instead. C class novae display characteristic cusps in their light curves, which appear as secondary additive on top of the base level S class type light curve shape \citep{Strope2010AJ}. Figure \ref{fig:cusp_nova} shows how there appears to be a strong additive component atop the expected S class curvature for V906 Car. Approximately 30 days after discovery, the nova reaches a peak rapidly at $m_V = 5.84 \pm 0.09$ before resuming to the decline law of S class light curve shape. 

\begin{figure}
\includegraphics[width=0.52\textwidth]{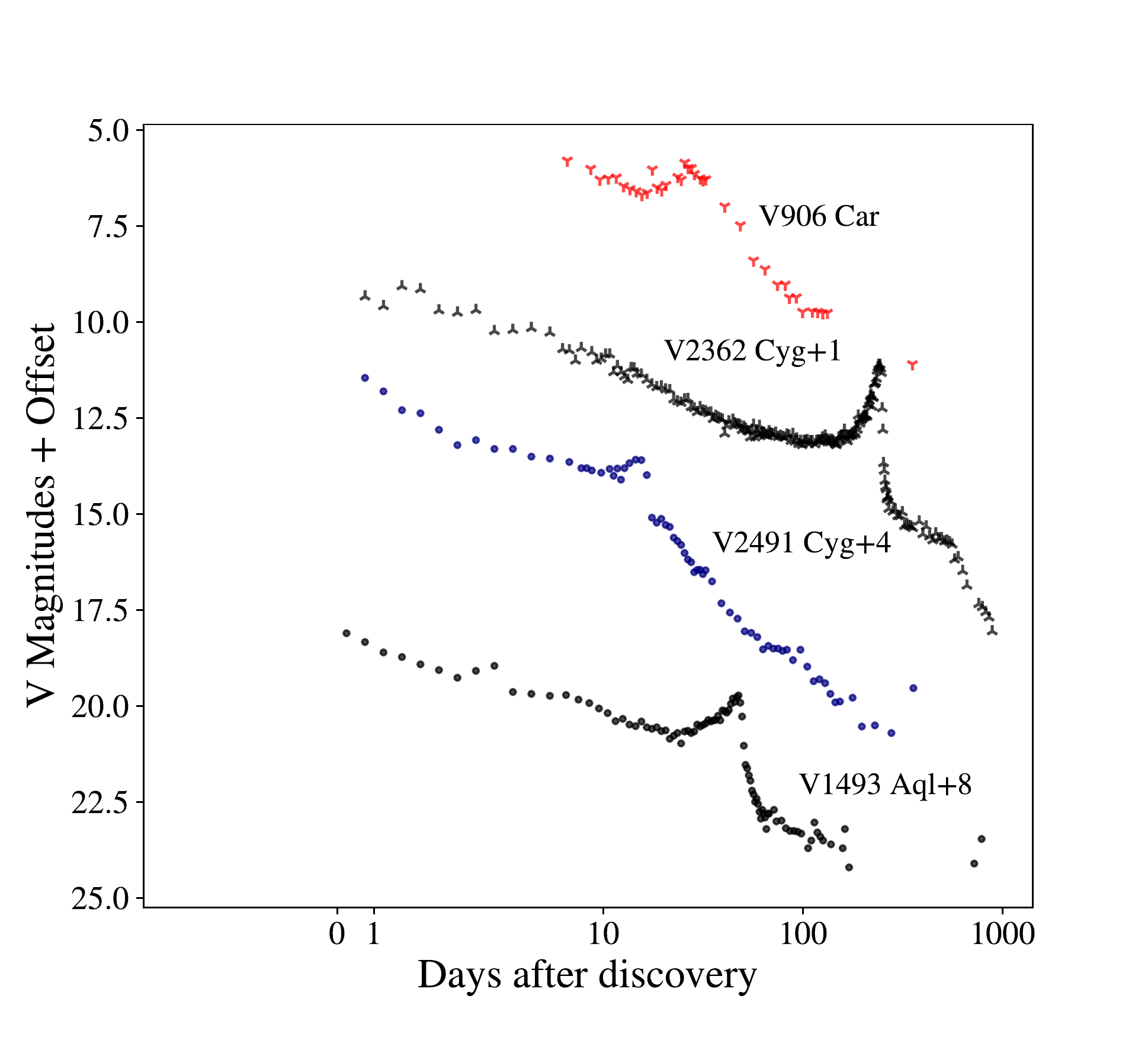}
\caption{Comparison of ASASSN-18fv with other C class novae in logarithmic time with LCO 1.0m $V$. \label{fig:class}}
\end{figure}

Our calculations of the decline rates (in units of magnitudes per logarithmic time) of V906 Car post primary peak yields the following results for each of its declining phases (see Table 6) show that its decline characteristics conform to declines observed in C class novae. Our calculations of the strength of the cusps (see Table 7) also show that V906 Car conforms to the morphology of a C class nova, with an increase from $V_{\rm{min}}$ to $V_{\rm{cusp}}$ of 0.78 mag and a significant decrease from $V_{\rm{cusp}}$ to $V_{\rm{base}}$ of 2.55 mag. Comparatively, V906 Car resembles V2491 Car the most, with similar decline patterns and smaller cusps relative to their general decline. V906 Car's smaller cusp compared to V2352 Cyg and V1492 Aql is consistent with the expectations by \citet{Strope2010AJ} that the later the cusp appears, the stronger it is. A comparison of V906 Car in $V$ band with other C class novae is presented in in Figure \ref{fig:class}.

\begin{table}
\begin{minipage}{1.1\linewidth}
\begin{small}
\caption{Slope comparison with other C class novae}
\centering
\begin{tabular}{cccc}
\hline
Nova & Slope 1  & Slope 2 & Slope 3    \\ \hline
 V1493 Aql$^a$ 	& $-4.0$ & $-2.0$ & ... \\
 V2362 Cyg$^a$ 	& $-2.7$ & $-2.7$ & $-10.8$ \\
 V2491 Cyg$^a$ 	& $-2.0$ & $-7$ & $-3.8$ \\
 V906 Car 	& $-2.24$ & $-3.31$ & $-3.94$ \\
\hline
\end{tabular}
\begin{tablenotes}
\item[\textdagger]$^a$ Figures presented are from \cite{Strope2010AJ}.
\end{tablenotes}
\label{tab:classcompare}
\end{small}
\end{minipage}
\end{table}

\begin{table}
\begin{minipage}{1.1\linewidth}
\begin{small}
\caption{Cusp comparison with other C class novae in V magnitude}
\centering
\begin{tabular}{cccc}
\hline
Nova & $V_{min}$  & $V_{cusp}$ & $V_{base}$    \\ \hline
 V1493 Aql$^a$ 	& $12.8$ & $11.7$ & $14.9$ \\
 V2362 Cyg$^a$ 	& $12.1$ & $10.1$ & $13.6$ \\
 V2491 Cyg$^a$ 	& $9.9$ & $9.6$ & $11.2$ \\
 V906 Car 	& $6.6$ & $5.8$ & $8.39$ \\
\hline
\end{tabular}
\begin{tablenotes}
\item[\textdagger]$^a$ Figures presented are from \cite{Strope2010AJ}.
\end{tablenotes}
\label{tab:cuspcompare}
\end{small}
\end{minipage}
\end{table}

\subsection{White Dwarf Mass} \label{subsec:wdmass}
Optical and infrared light curves of classical novae are approximately homologous among various white dwarf (WD) masses and chemical compositions when free-free emission from optically thin ejecta is spherical and dominates the continuum flux of novae \citep{2006ApJS..167...59H}. Such a homologous template light curve is called `a universal decline law'  \citep{2006ApJS..167...59H}. The timescale of the light curve depends strongly on the WD mass but weakly on the chemical composition, and hence we are able to roughly estimate the WD mass from the light-curve fitting. 

The template light curve for the universal law has a slope of flux $F \propto t_{\rm{init}}^{-1.75}$ in the middle from $t_2$ to $t_6$ below optical maximum, but it declines more steeply where  $F \propto t_{\rm{init}}^{-3.5}$ from $t_6$ to $t_{10}$ \citep{2006ApJS..167...59H}. The point of which the decline rate changes, caused by a quick decrease in the wind mass-loss rate, is known as the break time $t_{\rm{break}}$ caused by a quick decrease in the wind mass-loss rate \citep{2006ApJS..167...59H}. Once $t_{\rm{break}}$ is determined, we can derive the period of a UV burst phase, the duration of the optically thick wind phase, and the turnoff date of hydrogen shell-burning \citep{2006ApJS..167...59H}. 

As noted in \citet{2020NatAs.tmp...79A}, the X-ray emission from V906 Car has been heavily attenuated and reprocessed into optical and infrared light, both from absorption and X-ray suppression in corrugated shock fronts \citep{2015MNRAS.450.2739M, 2018MNRAS.479..687S}.  This makes it difficult to use the X-ray emission to constrain the WD properties, and available X-ray data reported in \citet{2020NatAs.tmp...79A} included a non-detection in the 0.3$-$10.0\,keV range with a 3$\sigma$ upper limit. To provide some constraints on the WD mass, we compared the late time optical emission with the grid of models in \citet{2006ApJS..167...59H}, recognizing the limitations and uncertainties inherent in using only optical data and possible systematic differences between the V906 Car nova from the novae used in calibrating the grid of models within \citet{2006ApJS..167...59H}.

From studying our observed optical photometry, we estimate that log\,$t_{\rm{break}}$ = 2.67, or $\sim$ 470 days. Using a grid of models of CO novae from the literature, we observe that our break time provides an upper limit to the WD mass of $M_{WD} < 0.8 M\textsubscript{\(\odot\)}$, for compositions with 0.35 $< X <$ 0.70, and CNO metallicity between $0.03< X_{\rm{CNO}}<0.3$ \citep{2006ApJS..167...59H}.

\section{Conclusion} \label{sec:conclusion}

We have presented optical and near-IR photometry of V906 Car. We found that the nova is a fast nova, peaking at $V_{max}=5.84 \pm 0.09$ mag and has rapid decline times of $t_{2,v}$ = 26.2 days and $t_{3,v}$ = 33.0 days. We have also presented pre-discovery data from Evryscope, showing that the nova is already visible at $g'\simeq13$ mag five days before discovery, and the early photometry provides rare data of evolution of this event well before the peak magnitude is observed. 

Our spectroscopic analysis yields a reddening of $E(B-V) \sim 0.36 \pm 0.04$ and an extinction of $A_V \sim 1.116^{+0.538}_{-0.385}$ mag in the line of sight to the nova, making it a likely member of the Carina Nebula. 

Our modelling of the nova progenitor in quiescence shows a good agreement with the emission coming from two black bodies. The hotter component has $T_{BB1} \sim 1.5 \times 10^4$\,K and $L_{BB1}\simeq 2.8 \times 10^{-1}$\,\Lsun. With a radius of $R_{BB1} \sim 50 \times 10^8$\,cm. this emission is likely associated with the accretion disk. The colder component represents the donor star. Its luminosity $L_{BB2} \sim 3\times 10^{-2}$\,\Lsun and radius $R_{BB2}\sim 1.7 \times 10^{10}$\,cm suggest that the star is a 0.23$-$0.42\,\Msun K-M type dwarf which is likely being heated by the WD companion.
When a steady accretion disk model is used to interpret the system in quiescence, the analysis yields a progenitor mass of $M_* =0.71_{-0.19}^{+0.23}$\,\Msun and radius of $R_{*} = 21.2_{-0.5}^{+0.5}\times 10^8$\,cm, consistent with a WD. 

An unambiguous classification for the nova is difficult, but overall the nova most closely resembles a C class nova, with a steep decline slope of $\alpha = -3.94$. The steep decline slope is consistent with other C class novae, such as V2362 Cyg, V2491 Cyg, and V1493 Aql. Our data also allows us to provide an additional constraint for the upper limit on the WD mass of $M_{WD} < 0.8 M\textsubscript{\(\odot\)}$. 

We have used combined infrared and optical data spanning a wide range of times from well before peak magnitude to months afterwards to characterize the nearby V906 Car nova and its progenitor system. Future work using complementary data in other wavelengths will help further refine the characteristics of this peculiar transient and help to improve current astrophysical models for these systems.

\acknowledgments

\begin{small}
B.E.P. would like to acknowledge the support from Soka University of America, and from funding from NSF grant (AST-1440341) and an NSF PIRE Grant 1545949.

N. B. would like to acknowledge that this work is part of the research programme VENI, with project number 016.192.277, which is (partly) financed by the Netherlands Organisation for Scientific Research (NWO).

B.B. is supported by the NSF grant AST-1812874.

Based on data products from observations made with ESO Telescopes at the La Silla Paranal Observatory under programme ID 177.D-3023, as part of the VST Photometric H{alpha} Survey of the Southern Galactic Plane and Bulge (VPHAS+, www.vphas.eu).

This work has made use of data from the European Space Agency (ESA) mission
{\it Gaia} (\url{https://www.cosmos.esa.int/gaia}), processed by the {\it Gaia}
Data Processing and Analysis Consortium (DPAC,
\url{https://www.cosmos.esa.int/web/gaia/dpac/consortium}). Funding for the DPAC
has been provided by national institutions, in particular the institutions
participating in the {\it Gaia} Multilateral Agreement.

This research has made use of the SVO Filter Profile Service (http://svo2.cab.inta-csic.es/theory/fps/) supported from the Spanish MINECO through grant AYA2017-84089.

\end{small}

%

\vspace{5mm}
\facilities{CTIO:1.3m, 1.5m, Evryscope, LCO: 1m}


\software{Astropy \citep{2013A&A...558A..33A}, Cyanogen Imaging MaxIm DL v6 (\url{http://diffractionlimited.com/product/maxim-dl/}), LCO BANZAI pipeline \citet{2018zndo...1257560M}, BBFit (\url{https://github.com/nblago/utils}), emcee \citep{2013PASP..125..306F}}

\bibliography{references}

\end{document}